\documentclass[runningheads,a4paper]{llncs}
\usepackage{amsmath}
\usepackage{amsfonts}
\usepackage{amssymb}
\usepackage{graphicx}
\usepackage[usenames,dvipsnames]{color}
\usepackage{url}
\urldef{\mailsc}\path|tthieu, rmelnik}@wlu.ca|    
\newcommand{\keywords}[1]{\par\addvspace\baselineskip
\noindent\keywordname\enspace\ignorespaces#1}

\begin{document}

\mainmatter  

\title{Effects of random inputs and short-term synaptic plasticity in a LIF conductance model for working memory applications}

\titlerunning{Effects of random inputs and short-term synaptic plasticity in a LIF model}

%
%
\author{Thi Kim Thoa Thieu \inst{1} %
	\and Roderick Melnik \inst{1,2}}
\authorrunning{Thoa Thieu and Roderick Melnik}

\institute{MS2Discovery Interdisciplinary Research Institute, Wilfrid Laurier University, \\75 University Ave W, Waterloo, Ontario, Canada N2L 3C5 
	\\
	\and
	BCAM - Basque Center for Applied Mathematics, Bilbao, Spain\\
	\mailsc}

%
%

\toctitle{Lecture Notes in Computer Science}
\tocauthor{Authors' Instructions}
\maketitle

\begin{abstract}
	
Working memory (WM) has been intensively used to enable the
temporary storing of information for
processing purposes, playing an important 
role in the execution of various cognitive tasks. Recent studies have shown that information in WM is not only maintained through persistent recurrent activity but also can be stored in activity-silent states such as in short-term synaptic plasticity (STSP). Motivated by important applications of the STSP mechanisms in WM, the main focus of the present work is on the analysis of the effects of random inputs on a leaky integrate-and-fire (LIF) synaptic conductance neuron under STSP. Furthermore, the irregularity of spike trains can carry the information about previous stimulation in a neuron. A LIF conductance neuron with multiple inputs and coefficient of variation (CV) of the inter-spike-interval (ISI) can bring an output decoded neuron. Our numerical results show that an increase in the standard deviations in the random input current and the random refractory period can lead to an increased irregularity of spike trains of the output neuron. 
\keywords{Working memory  $\cdot$ short-term synaptic plasticity $\cdot$  LIF $\cdot$ Langevin stochastic models $\cdot$  spike time irregularity $\cdot$ random input currents $\cdot$  synaptic conductances $\cdot$  neuron spiking activities $\cdot$  uncertainty factors $\cdot$  membrane and action potentials $\cdot$  neuron refractory periods}
\end{abstract}

\section{Introduction}

Human working memory (WM) is a crucial part of human brain studies. In general, the simplest assumption is that information in WM is maintained through persistent recurrent activity. However, recent studies have shown that information can be maintained without persistent firing, namely, information can be stored in activity-silent states. Short-term synaptic plasticity (STSP) is one of the candidate mechanisms for storing information in activity-silent states, STSP leads to rapid changes in the strength of connections between neurons that reflects new information being presented to the network system \cite{Mongillo2008,Pals2020}. STSP strongly affects the information processing in the nervous system. STSP is used to study by using the description of intracellular
recordings of postsynaptic potentials or currents evoked by presynaptic spikes. However, STSP can also affect the statistics of postsynaptic spikes \cite{Ghanbari2017}. A comprehensive description of the combined effect of both short-term facilitation
and depression on noise-induced memory degradation in one-dimensional continuous
attractor models has been provided in \cite{Seeholzer2019}. STSP makes neurons sensitive to the distribution of presynaptic population firing rates \cite{Tauffer2021}. On the other hand, the dynamics of firing rate and irregularity of single
neurons are closely connected \cite{Hamaguchi2011,Fitz2020}. Using a
computational model to study the formation of silent assemblies in a network of spiking neurons, the authors in \cite{Gallinaro2021} have found that even though the formed assemblies
were silent in terms of mean firing rate, they had an increased coefficient of variation of
inter-spike intervals.

In this paper, we consider the effects of random inputs to a  LIF conductance neuron with STSP for applications in WM. In particular, we develop a LIF synaptic conductance model under a facilitation type of short-term synaptic plasticity dynamics. We study the effects of random external current inputs and random refractory periods on the spiking activities of neurons in a cell membrane potential setting of such LIF conductance neuron. Our analysis is carried out by considering a Langevin stochastic dynamic in a numerical setting for a cell membrane potential with random inputs. The numerical results demonstrate that the random inputs affect the spiking activity of the neuron. Under a weak excitatory input to the LIF conductance neuron together with the short-term facilitation, the memory can be reactivated. Furthermore, an increase in the standard deviations of Gaussian white noise inputs can lead to an increase in the irregularity of spike trains of the output neuron. 

%
%

%
%

\section{Synaptic conductance model description}

The simplest assumption in the modelling of synapses is that the synaptic weights are fixed. To get closer to the real situation, we will investigate synapses whose weights change in some input conditions. One of the candidates for such changes in the synaptic weights is the STSP. In general, STSP is a phenomenon in which synaptic efficacy changes over time in a way that reflects the history of presynaptic activity. There are two types of STP: Short-Term Depression (STD) and Short-Term Facilitation (STF), with opposite effects on synaptic efficacy, which have been experimentally observed. 

The mathematical model of STSP is characterized by a limited pool of synaptic resources available for transmission $R$, which is the amount of available resources to the presynaptic neuron. For instance, the overall amount of synaptic vesicles at the presynaptic terminals. We know that the number of presynaptic resources changes in a dynamic fashion depending on the recent history of spikes. Specifically, at a presynaptic spike, the fraction $u$ (the fraction of resources used each time a neuron fires) of the available pool to be utilized increases due to spike-induced calcium influx to the presynaptic terminal. Then, $u$ is consumed to increase the post-synaptic conductance. During each spike, $u$ decays back to zero with time constant $\tau f$, while $R$ recovers to 1 with time constant $\tau d$. We define the following dynamics of excitatory (subscript $E$) STSP (see, e.g., \cite{Pals2020}):

{\large\begin{equation}\label{STP_1}
	\begin{cases}
	\frac{d u_E}{dt} &= -\frac{U_0 - u_E}{\tau_f} + U_0(1-u_E^{-})\delta (1-t_{sp}),\\
	\frac{d R_E}{dt} &= \frac{1 - R_E}{\tau_d} - u_E^{+}R_E^{-}\delta(1-t_{sp}),\\
	\frac{d g_E(t)}{dt} &= - \frac{g_E}{\tau_E} + \bar{g}_Eu_E^{+}R_E^{-}\delta(1-t_{sp}),
	\end{cases}
	\end{equation}}where $U_0$ is a constant determining the increment of $u$, $u_E^{-}$ and $R_E^{-}$ represent the corresponding values before the arriving spike, while $u_E^{+}$ denotes the moment right after the spike. In \eqref{STP_1}, $\bar{g}_E$ represents the maximum excitatory conductance, while $g_E(t)$ is calculated for all spike times $sp$. Here, $\delta(\cdot)$ denotes the Dirac delta function, while $\tau_E$ is the given time constant. Moreover, the dynamics of inhibitory STSP can be described by replacing the subscript $E$ with $I$ in system \eqref{STP_1}. 

\begin{figure}[h!]
	\centering
	\includegraphics[width=0.85\textwidth]{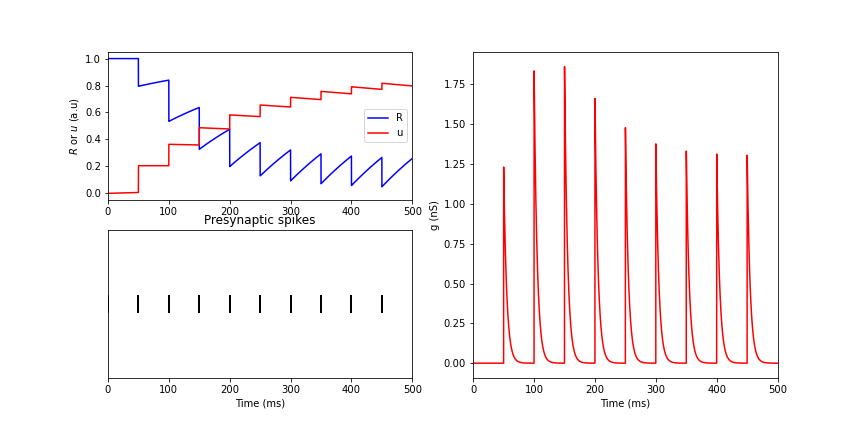}
	\caption{[Color online] Short-term synaptic facilitation (STF) changes for firing rates $r_i = r_e = 20$ of the presynaptic spike train and the amplitude synaptic conductance $g$ changes with every incoming spike until it reaches its stationary state.}
	\label{fig:0}
\end{figure}

\begin{figure}[h!]
	\centering
	\includegraphics[width=0.85\textwidth]{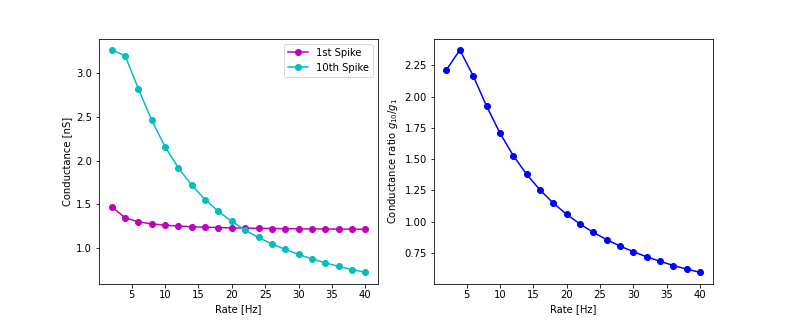}
	\caption{[Color online] The ratio of the synaptic conductance corresponding to the first and 10th spikes as a function of the presynaptic firing rate in the STF case.}
	\label{fig:1}
\end{figure}

 STSP involves mechanisms for both facilitation of transmitter
release, where synaptic strength increases with consecutive presynaptic spikes, and depression with synaptic strength decreases. The dynamics of $u$ and $R$ determine if the joint effect of $uR$ is dominated by depression or facilitation. In the regime of $\tau_d \gg \tau_f$ and for large $U_0$, the synapse is STD-dominated due to an initial spike incurs a large drop in $R$ that takes a long time to recover. In the regime of $\tau_d \ll \tau_f$ for small $U_0$, the synapse is STF-dominated since the synaptic efficacy is increased gradually by spikes. The kinetic dynamics of depressed and facilitated synapses observed in many cortical areas have been successfully reproduced by using such STSP phenomenological model.
In this work, we consider a LIF synaptic conductance model with STSP for working memory. This model is sustained by calcium-mediated synaptic facilitation in the recurrent connections of neocortical networks. The facilitating transmission is displayed by all excitatory-to-excitatory connections in the system. Moreover the amount of available resources $(R_E \text{ such that } 0 \leq R_E \leq 1)$ and the utilization parameter $u_E(x)$ modulate the synaptic efficacy. Such factors define the fraction of resources used by each spike, reflecting the residual calcium level. During a spike, the amount of $u_ER_E$ is used to produce the postsynaptic current, thus $R_E$ reduces. This process is known as neurotransmitter depletion \cite{Mongillo2008}. See, e.g., Figs. \ref{fig:0}-\ref{fig:1} for (STF) changes represented for firing rates of the presynaptic spike train. The amplitude synaptic conductance $g$ changes with every incoming spike until it reaches its stationary state, and the ratio of the synaptic conductance corresponding to the 1st and 10th spikes changes as a function of the presynaptic firing rate in the STF case. In Fig. 2, we observe that the small fluctuations are visible in the data presented for the conductance corresponding to the 10th spike and the conductance ratio of the synaptic conductance corresponding to the 1st and 10th spikes. Such small fluctuations come from the fact that total synaptic resources are finite and recover in a finite time. Hence, at high frequency inputs, synaptic resources are rapidly neglected at a higher rate than their recovery. After the first few spikes, only a small number of synaptic resources are left. Therefore, the steady-state synaptic conductance at high frequency inputs decreases.


\subsection{LIF synaptic conductance dynamics}

In this section, we consider a model of synaptic conductance dynamics under the STSP presented in \eqref{STP_1}. In particular, neurons receive myriad excitatory and inhibitory synaptic inputs at dendrites. To understand better the mechanisms underlying neuronal computation, we investigate the dynamics of STSP in a LIF neuron via electrophysiological recording techniques. 

In general, the synaptic input in vivo is characterized by the combination of excitatory neurotransmitters. Such excitatory neurotransmitters depolarize the cell and drive it towards a spike threshold. Inhibitory neurotransmitters hyperpolarize it, driving it away from the spike threshold. These factors cause specific ion channels on the postsynaptic neuron to open. Then, the neuron's conductance changes. Therefore, the current will flow in or out of the cell (see, e.g., \cite{Gerstner2014,Thieu2022}).
This synaptic conductance process can be modelled by assuming that the presynaptic neuron’s spiking activity produces transient changes in the postsynaptic neuron’s conductance $(g_{\text{syn}}(t))$. Such conductance transients $(g_{\text{syn}}(t))$ can be generated by using the system \eqref{STP_1}.
%

Using Ohm’s law allows us to convert conductance changes to the current as follows:

\begin{align}
	I_{\text{syn}}(t) = g_{\text{syn}}(t)(V(t) - E_{\text{syn}}), 
	\end{align}
where $E_{\text{syn}}$ denotes the direction of current flow of the excitatory ($E_E$) or inhibitory ($E_I$) of the synapse.

%

The total synaptic input current $I_{\text{syn}}$ is the combination of both excitatory and inhibitory inputs. Suppose the total excitatory and inhibitory conductances received at time $t$ are $g_E(t)$ and $g_I(t)$, and their corresponding reversal potentials are $E_E$ and $E_I$, respectively. The total synaptic current can be defined as (see, e.g., \cite{Li2019}): 

\begin{align}
	I_{\text{syn}}(V(t),t) = -g_E(t) (V-E_E) - g_I(t)(V-E_I). 
	\end{align}

Next, we note that the corresponding membrane potential dynamics of the LIF neuron under synaptic current can be described as follow (see, e.g., \cite{Li2019})

\begin{align}\label{main_eq}
	\tau_m \frac{d}{dt}V_m(t) &= -( V_m(t) - E_L)  - \frac{g_E(t)}{g_L}(V_m(t) - E_E) -  \frac{g_I(t)}{g_L}(V_m(t) - E_I) + \frac{I_{\text{inj}}}{g_L}, 
\end{align}
where $V_m$ is the membrane potential, $I_{\text{inj}}$ is the external input current, $\tau_m$ is the membrane time constant, $g_L$ denotes the leak conductance, while $E_L$  is the leak potential. 
%

We consider a random synaptic input by introducing the following random input current (additive noise)
$I_{\text{Inj}} = \mu_{\text{Inj}} + \sigma_{\text{Inj}} \eta(t)$ (pA), where $\eta$ is the Gaussian white noise with mean $\mu_{\text{Inj}}$ and standard deviation $\sigma_{\text{Inj}}$.

By considering such Gaussian white noise input currents, the equation \eqref{main_eq} can be considered as the following Langevin stochastic equation (see, e.g., \cite{Roberts2017}):

\begin{align}\label{main_eq02}
	\tau_m \frac{d}{dt}V_m(t) &= -( V_m(t) - E_L)  - \frac{g_E(t)}{g_L}(V_m(t) - E_E) -  \frac{g_I(t)}{g_L}(V_m(t) - E_I) \nonumber\\&+ \frac{1}{g_L}(\mu_{\text{Inj}} + \sigma_{\text{Inj}} \eta(t))
	 \quad \text{ if } V(t) \leq V_{\text{th}}. 
\end{align}

In this paper we investigate the effects of random refractory periods. We define the random refractory periods $t_{\text{ref}}$ with $t_{\text{ref}} = \mu_{\text{ref}} + \sigma_{\text{ref}}\mathcal{N}$, where $\mathcal{N}$ is the normal distribution. 

In our model, to approximate the stochastic neuronal firings, we use the simplest input spikes with the Poisson process \cite{Kuhn2004,Thieu2022}.   The input spikes will be added to the system via the quantity $\delta(t-t_{sp})$ in \eqref{STP_1}. We assume that the input spikes are given when every input spike arrives independently of other spikes. For designing a spike generator of spike train, we define the probability of firing a spike within a short interval  (see, e.g. \cite{Dayan2005}) $P(1 \text{ spike during } \Delta t) = r_{j}\Delta t$, where $j=e,i$ with $r_e, r_i$ representing the instantaneous excitatory and inhibitory firing rates, respectively. A Poisson spike train is generated by first subdividing time into a group of short intervals through small time steps $\Delta t$. At each time step, we define a random variable $x_{\text{rand}}$ with uniform distribution
over the range between 0 and 1. Then, we compare this quantity with the probability of firing a spike, which reads:

\begin{align}
	\begin{cases}
		r_j\Delta t > x_{\text{rand}}, \text{ generate a spike},\\
		r_j \Delta t \leq x_{\text{rand}}, \text{ no spike
			is generated}.
	\end{cases}
\end{align}

\subsection{Firing rate and spike time irregularity }

In general, the irregularity of spike trains can carry information about previous stimulation in a neuron. A LIF conductance neuron with multiple inputs and coefficient of variation (CV) of the inter-spike-interval (ISI) can bring an output decoded neuron. In particular, we have found that the increase of $\sigma_{\text{Inj}}$ and $\sigma_{\text{ref}}$ can lead to an increase in the irregularity of the spike trains (see also \cite{Gallinaro2021}).

Spike regularity can be calculated as the following coefficient of variation of the inter-spike-interval (see, e.g., \cite{Gallinaro2021}):
$$CV_{\text{ISI}} = \frac{\sigma_\text{ISI}}{\mu_\text{ISI}},$$
where $\sigma_\text{ISI}$ is the standard deviation and $\mu_\text{ISI}$ is the mean of the ISI of an individual neuron.

In the next section, we plot and analyze the output firing rate as a function of Gaussian white noise mean or direct current value, known as the input-output transfer function of the neuron.
 

%
%
\section{Numerical results for the LIF synaptic conductance model}\label{num}

In this subsection, we take a single pyramidal neuron at the dendrite and study how the neuron behaves under STF dynamics and when it is bombarded with both excitatory and inhibitory spike trains (see, e.g., \cite{Li2019,Mongillo2008,Pals2020}). 

In what follows, the simulations have been carried out by a modification of the numerical method provided in the open source framework at \url{https://github.com/} (see W2D3 Biological Neuron Models in the Neuromatch Academy directory).

In the simulations, we choose the parameter set as follows: $E_E = 70$ (mV), $E_L = -60$ (mV), $E_I = -10$ (mV), $V_{\text{th}} = -55$ (mV), $V_{\text{reset}} = -70$ (mV), $\Delta t = 0.1$, $\tau_m = 10$ (ms), $r_e = 20$, $r_i = 20$, $n_E = 20$ spikes, $n_I = 80$ spikes, $\bar{g}_E = 1.2\times 4$ (nS), $\bar{g}_I = 1.6\times 4$ (nS), $\tau_E = 5$ (ms), $\tau_I = 100$ (ms), $U_{\text{0E}} = U_{\text{0I}} = 0.2$, $\tau_{\text{dE}} = \tau_{\text{dI}} =  200$ (ms), $\tau_{\text{fE}} = \tau_{\text{fI}}= 1500$ (ms) . Here, $n_E$ and $n_I$ represent the number of excitatory and inhibitory presynaptic spike trains, respectively. These parameters have also been used in  for dynamic
clamp experiments and we take them for our model validation. In this section, we use the excitatory and inhibitory conductances provided in Fig. \ref{fig:1} for all of our simulations.
Further, we use the experimental data provided in \cite{Pals2020} and \cite{Li2019} for our model validation. 

\begin{figure}[h!]
	\centering
	\includegraphics[width=0.85\textwidth]{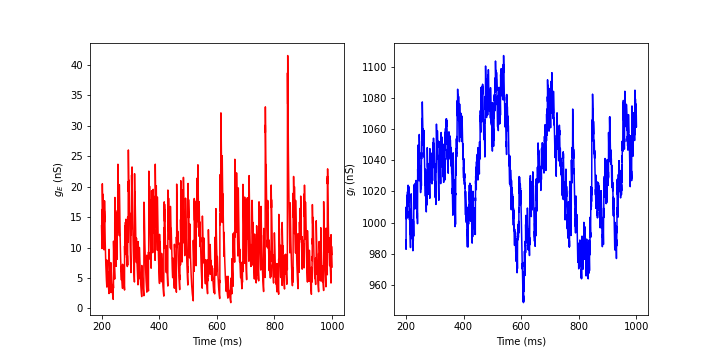}
	\caption{[Color online] Left: Excitatory conductances profile. Right: Inhibitory conductances profile.}
	\label{fig:2}
\end{figure}

\begin{figure}[h!]
	\centering
	\includegraphics[width=0.9\textwidth]{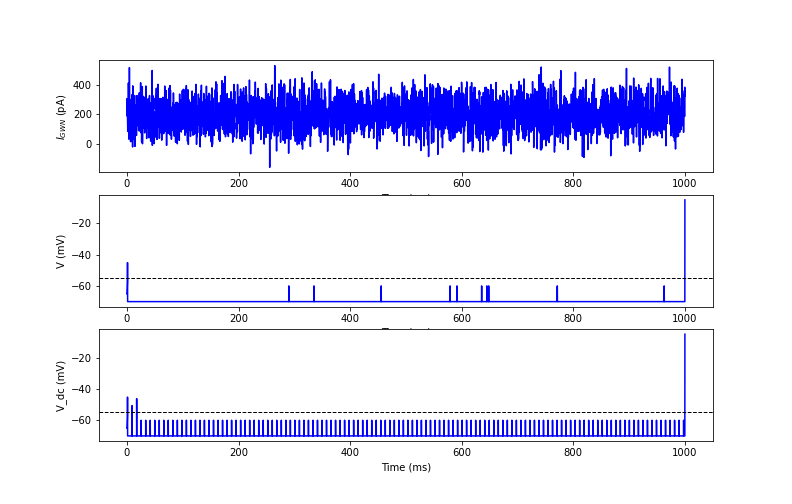}
	\caption{[Color online] Top row: Gaussian white noise current profile. Middle row: Time evolution of membrane potential $V(t)$ with Gaussian white noise current. Bottom row: Time evolution of membrane potential $V(t)$ with direct input.}
	\label{fig:3}
	
\end{figure}

\begin{figure}[h!]
	\centering
	\includegraphics[width=0.9\textwidth]{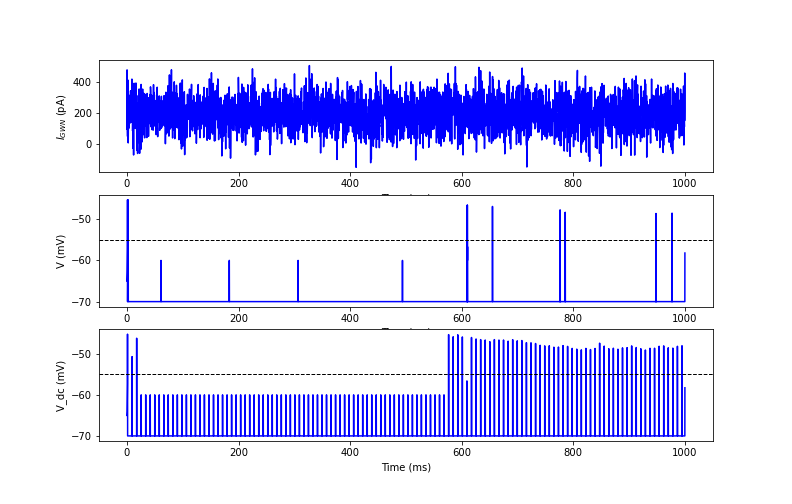}
	\caption{[Color online] Top row: Gaussian white noise current profile. Middle row: Time evolution of membrane potential $V(t)$ with Gaussian white noise current. Bottom row: Time evolution of membrane potential $V(t)$ with direct input.}
	\label{fig:3-2}
	
\end{figure}

\begin{figure}[h!]
	\centering
	\includegraphics[width=0.85\textwidth]{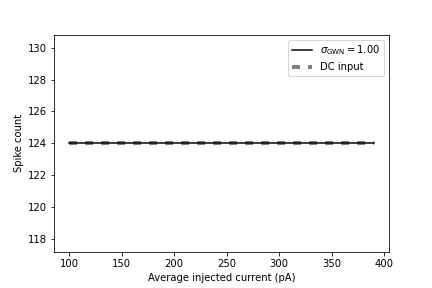}
	\caption{[Color online] The input-output transfer function of the neuron, output firing rate as a function of input mean. Parameters: direct input current  $\sigma_{\text{Inj}} = 1$, $t_{\text{ref}} = 8$ (ms).}
	\label{fig:4}
	
\end{figure}

\begin{figure}[h!]
	\centering
	\includegraphics[width=0.85\textwidth]{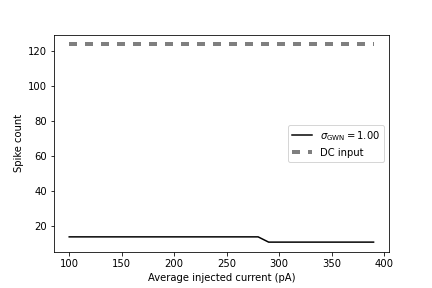}
	\caption{[Color online] The input-output transfer function of the neuron, output firing rate as a function of input mean. Parameters: $\sigma_{\text{Inj}} = 1$, $\mu_{\text{ref}} = 8$, $\sigma_{\text{ref}} = 1$, $t_\text{ref} = 8$ (ms) for direct input current.}
	\label{fig:5}
	
\end{figure}

\begin{figure}[h!]
	\centering
	\includegraphics[width=0.85\textwidth]{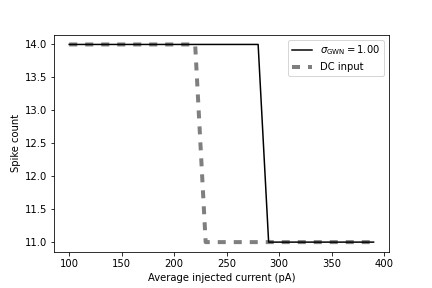}
	\caption{[Color online] The input-output transfer function of the neuron, output firing rate as a function of input mean. Parameters: $\sigma_{\text{Inj}} = 1$, $\mu_{\text{ref}} = 8$, $\sigma_{\text{ref}} = 1$.}
	\label{fig:6}
	
\end{figure}

\begin{figure}[h!]
	\centering
	\includegraphics[width=0.85\textwidth]{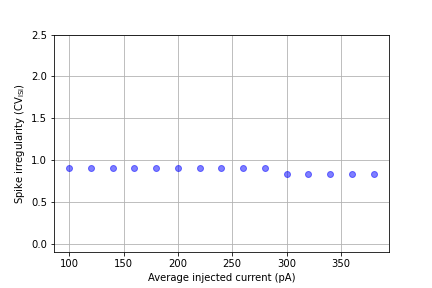}
	\caption{[Color online] Spike irregularity profile in the case with direct current. Parameters: $\sigma_{\text{Inj}} = 1$, $\mu_{\text{ref}} = 8$, $\sigma_{\text{ref}} = 1$.}
	\label{fig:7}
	
\end{figure}

\begin{figure}[h!]
	\centering
	\includegraphics[width=0.85\textwidth]{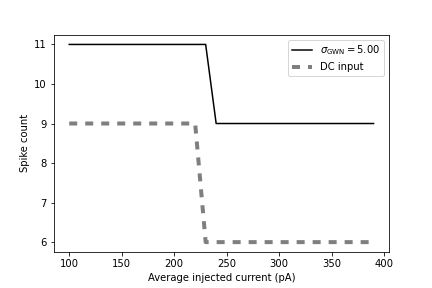}
	\caption{[Color online] The input-output transfer function of the neuron, output firing rate as a function of input mean. Parameters: $\sigma_{\text{Inj}} = 5$, $\mu_{\text{ref}} = 8$, $\sigma_{\text{ref}} = 2.5$.}
	\label{fig:8}
	
\end{figure}

\begin{figure}[h!]
	\centering
	\includegraphics[width=0.85\textwidth]{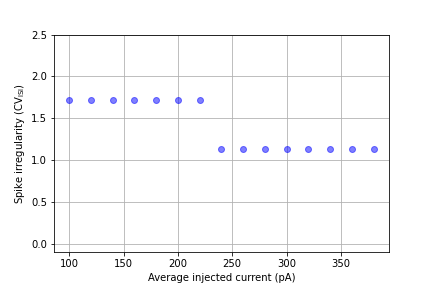}
	\caption{[Color online] Spike irregularity profile in the case with Gaussian white noise current. Parameters: $\sigma_{\text{Inj}} = 5$, $\mu_{\text{ref}} = 8$, $\sigma_{\text{ref}} = 2.5$.}
	\label{fig:9}
	
\end{figure}

The main numerical results of our analysis here are shown in Figs.\ref{fig:2}-\ref{fig:9}, where we have plotted the time evolution of the membrane potential calculated based on model \eqref{main_eq}, the input-output transfer function as well as the spike regularity profile of the neuron. We investigate the effects of random inputs on a LIF neuron under synaptic conductance dynamics and a facilitation type of short-term synaptic plasticity dynamics. By using a Poissonian spike input, we observe that the random external current and random refractory period influence the spiking activity of a neuron in the cell membrane potential. Our simulations demonstrate that as long as the synapses remain facilitated, the memory can be reactivated by presenting a weak excitatory input to the LIF conductance system, even though the neural activity is at the spontaneous level. Furthermore, the presence of random input current impacts the spiking activities of the system.

%

In particular, in Fig. \ref{fig:3}, we have plotted the Gaussian white noise current profile, the time evolution of the membrane potential $V(t)$ with Gaussian white noise input current and direct input current ($I_{\text{inj}} = I_{\text{dc}} = 200$ (pA)). In the case with Gaussian white noise input current in the second row of Fig. \ref{fig:3}, the neuron does not reach its threshold for a long time from 1 to nearly 1000 (ms). There are only the two spikes that come over the threshold. However, in the bottom row of Fig. \ref{fig:3}, with the direct input current, we observe that the neuron fires a spike within an interval of about 20 (ms) (see, e.g., \cite{Mongillo2008}). It is clear that the memory can be reactivated under a weak excitatory input to the LIF conductance system ($n_E = 20$ and $n_I = 80$ spikes). The presence of Gaussian white noise in the system increases the distance between each spike and decreases the spiking activity of the neuron compared with the case of direct input current. 

In Fig. \ref{fig:3-2}, by increasing the values of $\tau_{\text{dE}} = \tau_{\text{dI}} = 1400$ (ms), we observe that the spiking activity of the neuron increases in both two cases: Gaussian white noise input and direct input currents. In the second row of Fig. \ref{fig:3-2}, the spikes of the neuron increase compared to the cases presented in Fig. \ref{fig:3}. Specifically, in the third row of Fig. \ref{fig:3-2}, almost all spikes reach their threshold after a time of 570 (ms) in the case of direct input current. In the second row of Fig.\ref{fig:3-2}, the presence of the random input current in the system leads to an increase in the distance between spikes that decrease the spiking activity in the system. 

In Fig. \ref{fig:4}, we have plotted the spike count profile as a function of average injected current. With $\sigma_{\text{Inj}} = 1$ and $t_{\text{ref}} = 8$ (ms), we have 124 spikes for both cases: Gaussian white noise input and direct input currents. There is no difference in the spike count between the two cases.

In Fig. \ref{fig:5}, we consider the random refractory period for the case with Gaussian white noise current, and the standard refractory period $v$ for the case with the direct input current. We observe that the spike count dramatically reduces in the case of random input current and random refractory period compared to the cases in Fig. \ref{fig:4}. It is clear that the random refractory period affects the spiking activity of the system. 

In Fig. \ref{fig:6}, using random refractory periods for both cases: Gaussian white noise and direct input current, we observe that the spikes decrease in both cases. In particular, the spike count remains the same (14 spikes) from $I_{\text{Inj}} = 100$ (pA) to $I_{\text{Inj}} = 380$ (pA). Then it reduces to 11 spikes in the case of Gaussian white noise input current. Similarly, from $I_{\text{Inj}} = 225$ (pA), the spike count decreases from 9 spikes to 6 spikes also in the case of direct input current. This effect is caused by the presence of a random refractory period in the system. 

In Fig. \ref{fig:7}, we look at the corresponding spike irregularity profile of the spike count in Fig. \ref{fig:6}. We see that there is not much change in the coefficient of variation of the inter-spike-interval with values around 0.9. There is a slight decrease of the spike irregularity from the average injected current with values from 280 (pA) to 400 (pA).   

In Fig. \ref{fig:8}, we consider the same cases as in Fig. \ref{fig:6}. The only difference is that we increase the values of the standard deviations of the random input current and random refractory period to $\sigma_{\text{Inj}} = 5$ and $\sigma_{\text{ref}} = 2.5$. We observe that the spikes decrease when the average injected current increases. This is visible also in the corresponding spike irregularity profile in Fig. \ref{fig:9}, at the average injected current of value 250 (pA), we see a decrease of the spike irregularity coefficient $\text{CV}_{\text{ISI}}$ from 1.7 to 1.1. It is clear that even with a decrease in the spike irregularity the coefficient $\text{CV}_{\text{ISI}}$, in this case, is still larger than in the cases presented in Fig. \ref{fig:7}. This is due to the fact that when we increase the mean of the Gaussian white noise, at some point, the
effective input means are above the spike threshold and then the neuron operates in the so-called mean-driven regime. Hence, as the input is sufficiently high, the neuron is charged up to the spike threshold and then it is reset. This essentially gives an almost regular spiking.

Additionally, we notice that the presence of a random refractory period increases the distance of the time interval between two nearest neighbor spikes as well as decreases the spiking activity in the system. However, with a weak excitatory input to the LIF conductance system together with the STF, the spiking activity of the neuron still occurs and the memory can be reactivated. Under suitable values of average injected current as well as the values of random input current and random refractory period, the irregularity of spike trains increases. This effect leads to an improvement in the carrying of information about previous stimulation in the neuron.
\section{Conclusions}

We have proposed and described a LIF synaptic conductance model with random inputs. Using the description based on the Langevin stochastic dynamics together with the STSP, we have analyzed the effects of noise in a cell membrane potential. In particular, we have provided details of the model along with representative numerical examples. Our computational experiments have demonstrated that the presence of random input current and random refractory period decrease the spiking activity of the neuron in the system. The memory can be reactivated under a weak excitatory input to the LIF conductance system with STF. When the values of average injected current are large enough together with suitable values of the standard deviations of Gaussian white noise inputs, the irregularity of spike trains increases. A better understanding of uncertainty factors in LIF conductance neurons with STSP dynamics would contribute to further progress and model developments for WM in human brain studies.

\section*{Acknowledgments} 

 Authors are grateful to the NSERC and the CRC Program for their
support. RM is also acknowledging support of the BERC 2022-2025 program and Spanish Ministry of Science, Innovation and Universities through the Agencia Estatal de Investigacion (AEI) BCAM Severo Ochoa excellence accreditation SEV-2017-0718 and the Basque Government fund AI in BCAM EXP. 2019/00432.

\bibliographystyle{splncs}
\bibliography{mybibn}
\end{document}